\documentclass[final,authoryear,5p]{elsarticle}

\usepackage{graphicx}
\usepackage{amssymb}
\usepackage{amsthm}
\usepackage{amsmath}
\usepackage{tabularx}
\usepackage{nomencl}
\makenomenclature

\usepackage[
ps2pdf,%
a4paper=true,%
breaklinks=true,%
colorlinks=true,%
pdfauthor={Maciej Sznajder, Ulrich Geppert, Miroslaw Dudek},%
pdftitle={Degradation of metallic surfaces under space conditions, with particular emphasis on hydrogen recombination processes}%
]{hyperref}

\journal{Advances in Space Research}

\begin{document}

\begin{frontmatter}



\title{Degradation of metallic surfaces under space conditions, with particular emphasis on hydrogen recombination processes}


\author{Maciej Sznajder\corref{cor1}}
\ead{Maciej.Sznajder@dlr.de}
\address{DLR Institute for Space Systems, University of Bremen, Robert Hooke Str. 7, 28359 Bremen, Germany}

\author{Ulrich Geppert}
\address{DLR Institute for Space Systems, System Conditioning, Robert Hooke Str. 7, 28359 Bremen, Germany}
\address{Kepler Institute of Astronomy, University of Zielona G\'{o}ra, Lubuska 2, 65-265 Zielona G\'{o}ra, Poland}

\author{Miros\l{}aw Dudek}
\address{Institute of Physics, University of Zielona G\'{o}ra, Szafrana 4a, 65-069 Zielona G\'{o}ra, Poland}

\cortext[cor1]{Corresponding author}

\begin{abstract}
The widespread use of metallic structures in space technology brings risk of degradation which occurs under space conditions. New types of materials dedicated for space applications, that have been developed in the last decade, are in majority not well tested for different space mission scenarios. Very little is known how material degradation may affect the stability and functionality of space vehicles and devices during long term space missions.     

Our aim is to predict how the solar wind and electromagnetic radiation degrade metallic structures.  Therefore both experimental and theoretical studies of material degradation under space conditions have been performed. The studies are accomplished at German Aerospace Center (DLR) in Bremen (Germany) and University of Zielona G\'{o}ra (Poland).   

The paper presents the results of the theoretical part of those studies. It is proposed that metal bubbles filled with Hydrogen molecular gas, resulting from recombination of the metal free electrons and the solar protons, are formed on the irradiated surfaces. A thermodynamic model of bubble formation has been developed. 
We study the creation process of $\rm{H_2}$-bubbles as function of, inter alia, the metal temperature, proton dose and energy. Our model has been verified by irradiation experiments completed at the DLR facility in Bremen.

Consequences of the bubble formation are changes of the physical and thermo-optical properties of such degraded metals. We show that a high surface density of bubbles (up to $10^8$ $\rm{cm^{-2}}$) with a typical bubble diameter of $\sim 0.4$$\rm{\mu}$m will cause a significant increase of the metallic surface roughness. This may have serious consequences to any space mission.    

Changes in the thermo-optical properties of metallic foils are especially important for the solar sail propulsion technology because its efficiency depends on the effective momentum transfer from the solar photons onto the sail structure. This transfer is proportional to the reflectivity of a sail. Therefore, the propulsion abilities of sail material will be affected by the growing population of the molecular Hydrogen bubbles on metallic foil surfaces.        

\end{abstract}


\end{frontmatter}


\nomenclature{$\alpha$}{model parameters which determines gradient of bubble growth}
\nomenclature{$\alpha_{\rm{S}}$}{solar absorptance}
\nomenclature{$\alpha_{\rm{C}}$}{parameter of Lifshitz - Slyozov - Wagner theory}
\nomenclature{$A$}{area of the sample}
\nomenclature{$A_{\rm{b}}$}{area of the sample covered by the bubbles}
\nomenclature{$A_{\rm{t}}$, $A_{\rm{t}}^*$}{transition amplitude, complex conjugate of transition amplitude}
\nomenclature{$b$}{impact factor}
\nomenclature{$BS$}{backscatter coefficient}
\nomenclature{$d_{\rm{PR}}$}{projected range}
\nomenclature{$D_{H}$}{diffusion coefficient for H atoms in a given material}
\nomenclature[strings]{$\Delta(t)$}{difference of concentration of Hydrogen atoms at the bubble boundary and $C_{\infty}$}
\nomenclature{$C_{\infty}$}{concentration of Hydrogen atoms far beyond the bubble}
\nomenclature{$E$}{energy of incident ion}
\nomenclature{$E_{\rm{int}}$}{internal energy of molecules/atoms located at certain positions in the metal lattice}
\nomenclature{$E_{\rm{Y}}$}{Young module}
\nomenclature{$E_{\rm{min}}$}{ion's lowest energy recorded by the SOHO/ACE detector system}
\nomenclature{$E_{\rm{C}}$}{critical energy of incident ions above which ions pass through the material}
\nomenclature{$F_{\rm{gas,i}}$}{free energy of gas inside the $\rm{i^{th}}$ bubble}
\nomenclature{$F_{\rm{H}}$}{free energy of H atoms located outside bubbles in the metal lattice}
\nomenclature{$F_{\rm{H_2}}$}{free energy of $\rm{H_2}$ molecules located outside bubbles in the metal lattice}
\nomenclature{$F_{\rm{md,i}}$}{free energy of metal deformation caused by expanding $\rm{i^{th}}$ bubble}
\nomenclature{$F_{\rm{surf,i}}$}{surface free energy of the $\rm{i^{th}}$ bubble cap}
\nomenclature{$F_{\rm{config}}$}{free energy of a sample covered by bubbles}
\nomenclature{$\gamma$}{Poisson coefficient}
\nomenclature{$G_{\rm{i}}$}{fraction of $\rm{H_2}$ molecules merged into the $\rm{i^{th}}$ bubble}
\nomenclature{$H_{\rm{i}}$}{sum of partial derivatives}
\nomenclature{$H_{\rm{Sun}}(d)$}{radiation energy received from the Sun per unit area for a given distance $d$}
\nomenclature{$\eta_{\rm{max}}$}{relation between the number of $\rm{H_2}$ molecules and the H atoms in the metal lattice}
\nomenclature{$\epsilon_{\rm{H}}$}{migration energy of the H atom in the metal lattice}
\nomenclature{$\epsilon_{\rm{H_2}}$}{binding energy of the $\rm{H_2}$ molecule to a vacancy}
\nomenclature{$\epsilon_{\rm{cell}}$}{size of a grid cell that covers the irradiated sample}
\nomenclature{$\epsilon_{\rm{t}}$}{normal emittance}
\nomenclature{$I$}{proton flux}
\nomenclature{$I_{\rm{E}}$}{integrated proton flux}
\nomenclature{$I_{\rm{d}}$}{flux of protons at distance $d$ from the Sun}
\nomenclature{$k_{\rm{B}}$}{Boltzmann constant}
\nomenclature{$M_{\rm{u}}$}{molar mass of the sample's material}
\nomenclature{$N$}{number of time steps up to a given state of bubble growth}
\nomenclature{$N_{\rm{0}}$}{number of lattice sites}
\nomenclature{$N_{\rm{B}}^{\rm{T}}$}{total number of bubbles at the irradiated sample}
\nomenclature{$N_{\rm{cell}}$}{number of cells}
\nomenclature{$N_{\rm{H}}^{T}$}{total number of H atoms in the sample}
\nomenclature{$N_{\rm{H_2}}^{T}$}{total number of $\rm{H_2}$ molecules in the sample}
\nomenclature{$N_{\rm{H_2,i,j}}$}{number of $\rm{H_2}$ molecules add to the $\rm{i^{th}}$ bubble in the $\rm{j^{th}}$ time step}
\nomenclature{$N_{\rm{H_2}}^{\rm{out. bubbles}}$}{number of $\rm{H_2}$ molecules located in the lattice but outside bubbles}
\nomenclature{$N_{\rm{H,j}}$}{number of recombined H atoms in the $\rm{j^{th}}$ time step}
\nomenclature{$N_{\rm{diff,j}}$}{number of H atoms which diffuse from the sample out in the $\rm{j^{th}}$ time step}
\nomenclature{$\Omega$}{number of ways in which the $\rm{H_2}$ molecules can be arranged on the lattice sites}
\nomenclature{$Q$, $Q'$}{eigenstate of the incident ion and an electron before ($Q$) and after ($Q'$) recombination event}
\nomenclature{$P$}{probability of a recombination event}
\nomenclature{$p_{\rm{i}}$}{pressure of the gas inside the $\rm{i^{th}}$ bubble}
\nomenclature{$q$}{momentum}
\nomenclature{${\Delta}q_{\rm{i}}$}{momentum transfer of a photon to the $\rm{i^{th}}$ cell of the degraded foil}
\nomenclature{${\Delta}q_{\rm{max,i}}$}{momentum transfer of a photon to the $\rm{i^{th}}$ cell of a perfect mirror}
\nomenclature{$SC$}{Solar Constant}
\nomenclature{$d$}{distance from the Sun}
\nomenclature{$r_{i}$}{radius of the $\rm{i^{th}}$ bubble}
\nomenclature{$\bar{r}$}{average bubble radius}
\nomenclature{$r_{\rm{i,0}}$}{initial radius of the bubble}
\nomenclature{$r_{\rm{max,i}}$}{maximum radius of the $i^{th}$ bubble}
\nomenclature{${\Delta}R$}{decrease of the specular reflectivity}
\nomenclature{$S$}{entropy}
\nomenclature{$\sigma$}{surface tension}
\nomenclature{$\sigma_{\rm{SB}}$}{Stefan-Boltzmann constant}
\nomenclature{$\Sigma_{\rm{A}}$}{cross section for Auger recombination process}
\nomenclature{$\Sigma_{\rm{R}}$}{cross section for resonant recombination process}
\nomenclature{$\Sigma_{\rm{OBK}}$}{cross section for Oppenheimer-Brinkman-Kramers recombination process}
\nomenclature{$\Sigma_{\rm{total}}$}{total cross section for recombination processes}
\nomenclature{${\Delta}t_{\rm{j}}$}{time step}
\nomenclature{$T$}{temperature}
\nomenclature{$t_{\rm{s}}$}{number of days in space until a probe will collect a given dose of protons}
\nomenclature{$V_{\rm{i}}$}{volume of the $\rm{i^{th}}$ bubble}
\nomenclature{$V_{\rm{max,i}}$}{maximum volume of the $\rm{i^{th}}$ bubble}
\nomenclature{$V_{\rm{min}}$}{minimum volume of a bubble}
\nomenclature{$\xi$}{relation between the number of $\rm{H_2}$ molecules inside and outside the bubbles}
\nomenclature{$\Xi_{\rm{i,j}}$}{increment of increase of the $\rm{i^{th}}$ bubble radius during the $\rm{j^{th}}$ period of time}
\nomenclature{$\zeta_{\rm{H,j}}$}{number density of H atoms which may diffuse through the lattice in the $\rm{j^{th}}$ period of time}
\printnomenclature


\section{Introduction}
\label{introduction}

Metallic structures are commonly used in space technology. They build skeletons of spacecrafts, they protect satellites' interiors from rapid temperature changes (MLI blankets), or they are used as highly reflecting mirrors of optical space telescopes. Nowadays, metals are also used as thin layers on polyimide-type foils which have a broad usage in the solar sail technology. 

A failure of a space mission may be the result of a change of metallic structure properties caused by the environmental effects. Therefore, all of the materials planned for space applications have to be evaluated for their behavior under particle and electromagnetic radiation \citep{ecss-06C, astm-e512}. It is known from many of these evaluation tests that particle and electromagnetic radiation can significantly degrade materials \citep[see e.g.][]{lura, helt, sharma}.

Many kinds of metals have found their application in space industry, e.g. Aluminum, Copper, Nickel, Titanium, steels, and others \citep{ecss-71A,ecss-71C}. For instance Aluminum is one of the basic building materials of existing spacecrafts and is component of many subsystems. Copper is used in electrical, electronic and in general engineering applications. Nickel has its application e.g. in heating elements. Titanium is chosen in space applications for its mechanical-, temperature - properties, and chemical resistance \citep{ecss-71A,ecss-71C}. \newline However, it is very clearly stated that the radiation at the level existing in space does not modify the properties of metals \citep{ecss-71A}.

The here presented paper proves that the thin metallic foils are especially sensitive to the ion irradiation. The free electrons within the metals can in a well defined energy range recombine with the solar wind protons into neutral Hydrogen atoms. Recombination processes and their consequences have direct effects onto the foils' physical and thermo-optical properties \citep{sznajder}.

The result of a recombination of metal free electrons and solar protons is a formation of bubbles filled with Hydrogen molecular gas \citep[see e.g.][]{milacek, sznajder}. Bubble formation is one of the four degradation mechanisms caused by Hydrogen (referred to as embrittlement): formation of a hydride phase, enhanced local plasticity, grain boundary weakening and bubble formation \citep{myers, lu}. 

When $\rm{H_2}$-bubbles are formed on irradiated metallic surfaces, their reflectivity decreases with increasing surface density of the bubbles \citep{sznajder}. A reflectivity is a key parameter for many thin metallic foil applications e.g. in the solar-sail propulsion technology. 

Since the reflectivity is directly proportional to the momentum transfer from solar photons to the sail's material, its decrease will reduce the propulsion performance of a sail-craft just linearly proportional to the reduction of the reflectivity \citep{sznajder}. 

The future solar sail missions will be realized in the interplanetary medium \citep[see e.g.][]{gossamer, kawa, macdonald}. Unfortunately, the real degradation behavior of metallic samples is to a great extent unknown. Hence, detailed studies, both theoretical and experimental, are performed at German Aerospace Center (DLR) in Bremen, Germany with a cooperation of University of Zielona G\'{o}ra, Poland. 

The here presented thermodynamic model simulates the growth of $\rm{H_2}$-bubbles. The model input parameters are: the energy and flux of solar protons, type, and the temperature of the irradiated metal. The diffusivity of H in the metal lattice was taken into account, as well as back scattering effect (BS) of the solar protons irradiating the target. The model output is the velocity of bubble radius growth, the maximum possible bubble radius, and, for a given bubble density and average bubble radius, the reduction factor of the reflectivity with respect to its ideal value.

The paper is organized as follows. In Section \ref{recombination}, the typical proton flux spectra at the distance of $1.0$ AU from the Sun are presented. The value of the proton flux can be transformed to any distance orbit from the Sun, see Eq. \ref{flux_r}. Four recombination processes of the metal free electrons and the incident protons are presented: Auger -, resonant-, Oppenheimer-Brinkman-Kramers (OBK), and Radiative Electron Capture (REC) - process. In Section \ref{models} general principles and conditions for the $\rm{H_2}$-bubble formation are given. Next, the thermodynamic model of bubble formation is introduced. Also a simple model of specular reflectivity reduction due to growing population of bubbles is presented. In Section \ref{results} experimental results as well as validation of the thermodynamic model are presented. The reflectivity of metallic surfaces which are populated by a given surface density of bubbles is studied as well. Finally, in Section \ref{conclusions}, the conclusions are drawn.

\section{Recombination of protons into neutral Hydrogen atoms}
\label{recombination}

Devices, while operating in the interplanetary space, are exposed to solar wind and electromagnetic radiation. The solar wind, as the Sun's corona, is essentially made up of electrons and protons plus a small proportion of heavier ions, and it carries a magnetic field. Particles and fields are intimately coupled in plasmas \citep[][Ch.1]{sw}.

Extraterrestrial Sun observatories measure a few key solar wind parameters, e.g.: components of proton and electron velocity, their mean number density as well as components of the magnetic field. 

\begin{figure}[!h]
  \begin{center}
    \includegraphics[width=0.5\textwidth]{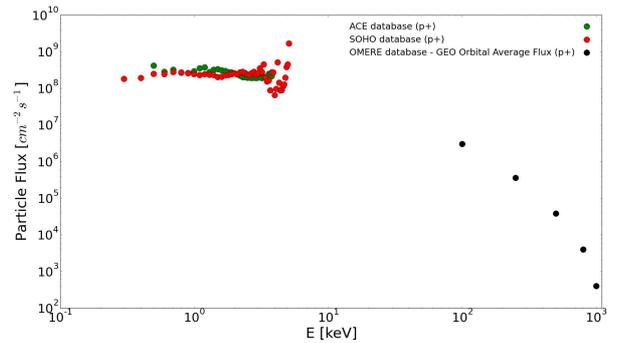}
  \end{center}
  \caption{Flux of solar protons as a function of energy. Data are taken from the SOHO, ACE, and OMERE database.}
  \label{p_flux}
\end{figure}

Fig. \ref{p_flux} shows solar proton at 1 AU distance from the Sun for its average activity. Proton fluxes are calculated by use of the data collected by the SOHO (since 1995) and the ACE (since 1997) satellites. The OMERE database is also considered.

When a probe is irradiated in space, it collects incident ions from a wide energy range. The range depends on the type and the thickness of an irradiated material. The thiner the target material, the less ions stuck in it. Therefore, there must exist a critical energy of incident ions ($E_{\rm{c}}$) above which they pass through the material. Hence, the integrated proton flux over the energies is:

\begin{equation}
  I_{\rm{E}} = \sum_{E_{\rm{min}}}^{E_{\rm{c}}} I(E),
\end{equation}  

\noindent where $E_{\rm{min}}$ is the ion's lowest energy recorded by the satellite's detector system. The $I_{\rm{E}}$ values are presented in the Table \ref{integrated_proton_flux}. To calculate the fluxes the ACE database was used. 

Obviously, the $E_{\rm{c}}$ depends on materials' type and thickness. The $7.5$ $\rm{\mu}$m $\rm{Upilex-S^{\textregistered}}$ foil covered on both sides with $100$ nm Al layer has been used to perform the first proton irradiation tests at DLR. For that coating thickness all of the protons with energy lower than $8$ keV will stuck in the Aluminum layer. Therefore, the integrated proton flux is $1.15 \times 10^{13}$ [$\rm{p^+}$$\rm{cm^{-2}}$$\rm{s^{-1}}$].    

\begin{table}[!ht]
\centering 
\caption{Integrated proton fluxes over the energies for $1$ AU distance orbit from the Sun.}
\vspace{3.pt}
\begin{tabular}{cc}
\hline
$E_{\rm{c}}$ [keV] & $I_{\rm{E}} \times 10^{13}$ [$\rm{p^+}$$\rm{cm^{-2}}$$\rm{s^{-1}}$]  \\
\hline
1.0 & 0.44 \\
1.5 & 0.68 \\ 
2.0 & 0.91 \\
2.5 & 1.06 \\
3.0 & 1.12 \\
4.0 & 1.14 \\
5.0 & 1.15 \\
9.0 & 1.15 \\
\hline
\end{tabular}
\label{integrated_proton_flux}
\end{table}

To estimate the flux of solar protons $I_{\rm{d}}$ at distance $d$ from the Sun, the following relation can be used:

\begin{equation} \label{flux_r}
  4\pi (1\rm{AU})^2 \times I_{\rm{E}} = 4\pi d^2 \times I_{\rm{d}}.
\end{equation}

Incident protons, while penetrating the metallic target, recombine with its free electrons to neutral Hydrogen atoms. There are four recombination processes of ions into neutral atoms:

\begin{enumerate}

\item Auger process

 In the Auger processes, an electron is captured by the incident ion to a bound state (here a proton) forming neutral Hydrogen. In order to conserve energy an Auger-electron or photon is released \citep[see e.g.][]{guinea, sols1, sols, echen, penalba, rosler, pauly}.

\item Resonant process

The resonant recombination proceeds when the incident ion is neutralized by an electron which is tunneled to a metastable state \citep{hag}. The inverse process is also possible. An electron which is in a metastable state with respect to the metallic ion can populate one of the free electron states of the metal only if it becomes free (the Pauli exclusion principle).  

The effect comes from the crystal structure itself. The resonant processes are due to the potential seen by the moving ion i.e. they are described in a frame where the incident ion is at rest \citep{sols}. From the point of view of the ion, there appear a moving periodic potential which gives rise to transitions between bound states of the ion-electron pair (composite) and free electron states \citep{echen}.

\item Oppenheimer-Brinkman-Kramers (OBK) process,

The OBK process is a capture process, where an inner or outer shell electron of a target atom is transferred to the moving ion \citep{sols}. In the literature there are many physical approaches \citep[see e.g.][]{chew, bran, lapicki, lin, belkic, ford, alston, miraglia, ghosh, grav, decker, datta}. Different results may be obtained depending on the approximation applied to the wave functions and the energy levels involved in the process \citep{echen}. 

In the OBK process the outer-shell electrons of metal ions experience a strong Coulomb field of the incident ion. The wave function of the electrons is distorted \citep{kuang, kuang1}. For the inner-shell capture, the screening effect of the outer-shell electrons of the metallic ions reduces the capture probability of an electron by the incident particle \citep{winter2, banyard, kuang}. 

\item Radiative Electron Capture (REC) process.

In the REC process an electron is transferred from the target atom to the incident ion with the simultaneous emission of a photon, see \citep{rec} or \citep[][Ch. 10, p. 151]{eich}. It occurs at ion energies exceeding $150$ MeV (for Aluminum as target material) \citep{rais}.

\end{enumerate}

Since the solar wind consists mainly of low ($\le 100$ keV) energetic protons, only the first three processes are responsible for recombination. 

The efficiency of the recombination processes is determined by their cross sections $\Sigma$. The cross section of each recombination process is calculated by use of the concept of so-called \textit{transition amplitude} $A_{\rm{t}}$. It determines the probability, $P=A_{\rm{t}}^*A_{\rm{t}}$, for a transition from an eigenstate $Q$ to $Q'$ \citep[][Ch. 15, p. 227]{penrose}. Here $P$ stands for probability of a recombination event i.e. that an electron is bound to an incident ion. $A_{\rm{t}}^*$ is the complex conjugate of A, $Q$ is an eigenstate of the incident ion and an electron before recombination takes place, while $Q'$ is an eigenstate of the ion-electron composite. Hence, the cross section of recombination event can be written as:

\begin{equation}
	\Sigma \sim \int_{0}^{\infty} A_{\rm{t}}^*A_{\rm{t}} \ b \ db,
\end{equation}

\noindent where $b$ is the impact parameter.

\begin{figure}[!h]
  \begin{center}
    \includegraphics[width=0.45\textwidth]{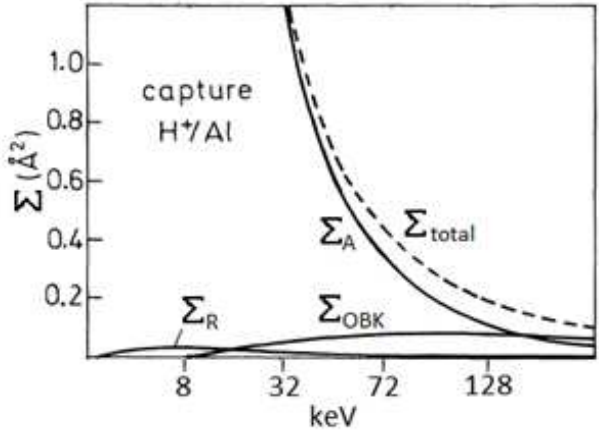}
  \end{center}
  \caption{Cross sections for capture processes of $\rm{H^{+}}$ ion: Auger ($\Sigma_{\rm{A}}$), resonant ($\Sigma_{\rm{R}}$) and OBK ($\Sigma_{\rm{OBK}}$) process. The dashed line represents the total cross section $\Sigma_{\rm{total}}$ of all processes \citep{sols1}.}
  \label{summary}
\end{figure}

Fig. \ref{summary} shows the different cross sections for capture processes when Aluminum is irradiated with protons ($\rm{H^+/Al}$) as a function of proton kinetic energy given in keV \citep{sols1}. One can see that the Auger process $\Sigma_{\rm{A}}$ is the dominant one. The resonant process $\Sigma_{\rm{R}}$ has negligible contribution to the total cross section. The OBK process $\Sigma_{\rm{OBK}}$ (in the literature it is often called the shell process) gives the main contribution to the total cross section at some MeV, depending on the target material \citep{rais}.

\section{Formation of molecular Hydrogen bubbles}
\label{models}

$\rm{H_2}$-bubbles occur as an irradiation damage. They change the physical properties of the irradiated surface and increase the erosion rate \citep{astrelin}. $\rm{H_2}$-bubbles are metal pockets filled with Hydrogen molecular gas. The tendency to form bubbles depends on the proton energy, integrated proton flux (dose), temperature of the target, crystallographic orientation of the irradiated surface as well as on impurities and defects in the sample \citep{daniels}. It is known from terrestrial laboratory experiments that for Aluminum the minimum dose of protons above which the process occurs is $\sim 10^{16}$ $\rm{H^+}$ $\rm{cm^{-2}}$ \citep[e.g.][]{milacek}. The temperature range in which bubbles were observed is between $288$ and $573$ K \citep[e.g.][]{milacek, daniels}. 

The procedure that was used to estimate the critical temperature ($573$ K) above which the process of bubble formation was stopped due to the bubble cracking mechanism was as follows. The Aluminum target was irradiated by a flux of protons at room temperature. When irradiation of the sample was stopped, the probe was heated up to higher temperatures. A significant increase of both, the surface density and sizes of the bubbles has been observed until the critical temperature was reached. That procedure, used by the authors, allows to capture more Hydrogen by the vacancies since during the irradiation, and at room temperature the vacancies will collect more Hydrogen than at elevated temperatures. The vacancy is a missing ion or point defect in metal lattice \citep{damask}. Also a diffusion of Hydrogen in Aluminum at room temperature is much lower than at temperatures reaching $\sim570$ K \citep{lind}. In space a probe is bombarded by the solar protons at the temperature which is related to its orbit. Therefore, the procedure presented by \citep{milacek, daniels} does not match the bubble formation mechanism under real space conditions.   

Hydrogen atoms are much smaller than metal ions, but they can introduce strain in a metal lattice when being absorbed as interstitial ions \citep{metzger, thomas, ren}. They can also change the electronic structure of near neighbor metal ions \citep{ren}. That causes an increase of the lattice energy. It may be decreased by the aggregation of the interstitial Hydrogen atoms into Hydrogen atom clusters, and then molecular Hydrogen bubbles \citep{ren}.  Hydrogen could not agglomerate into $\rm{H_2}$-clusters without the presence of vacancies. For instance a single vacancy in Aluminum can trap up to twelve H atoms. For comparison, a vacancy in Iron can trap only up to six H atoms \citep{lu}. Hydrogen atoms which do not constitute the $\rm{H_2}$ molecules and which are not trapped by the vacancies, diffuse through the metal lattice.

Molecular Hydrogen bubbles were observed also on different materials than Aluminum. Copper, Tungsten, Palladium and Iron were investigated \citep[e.g.][]{astrelin}. Bubbles are not forming on Tantalum and Vanadium. These metals are well known as \textit{blistering-resistant} materials \citep{astrelin}. However, they are not suitable for space applications where the surface reflectivity plays a crucial role, e.g. in the solar sail propulsion technology since their reflectance is $\sim50 \%$ lower than that of Aluminum \citep{optical}.   
    
\subsection{Formation of bubbles under space conditions}

Growth of molecular Hydrogen bubbles will be possible in the interplanetary space if the criterion of the minimum dose of protons is fulfilled. The temperature of the sample has to be high enough to start the bubble formation, but not too high to lose Hydrogen much too rapidly due to the high diffusivity of Hydrogen in metals.
 
Under the simplifying assumption that the Sun generates only mono-energetic 5 keV protons, the criterion of minimum dose of protons will be fulfilled after 116 days for 1 AU distance orbit from the Sun. Obviously, taking into account proton fluxes from the whole energy range, the criterion will be fulfilled much earlier.

The temperature of a foil placed in a given distance $d$ from the Sun can be calculated from the balance of heating and cooling by:

\begin{equation}
T = \left( \frac{A_{\rm{a}}}{A_{\rm{e}}} \frac{\alpha_{\rm{S}}}{\epsilon_{\rm{t}}} \frac{H_{\rm{Sun}}}{\sigma_{\rm{SB}}} \right)^{\frac{1}{4}}, \qquad H_{\rm{Sun}} = \frac{1 \ \rm{SC}}{d^2}.
\end{equation}

\begin{figure}
\centering
\includegraphics[width=0.5\textwidth]{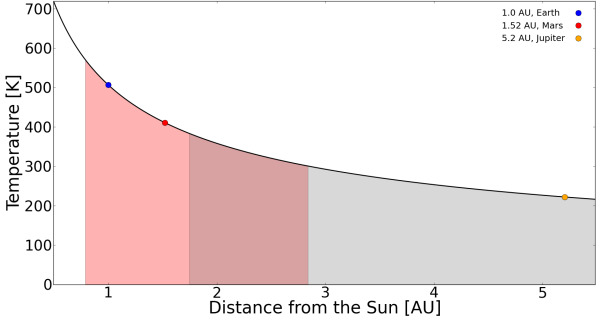}
\caption{Temperature of the $\rm{Upilex-S^{\textregistered}}$ foil covered on both sides with $100$ nm vacuum deposited Aluminum layer as a function of the distance from the Sun. The red area represents temperature range in which the bubble formation was reported in the literature. The dark-red area is the temperature range in which the formation has been confirmed by studies performed at DLR.}
\label{UBE_temperature}
\end{figure}

\noindent Here, $A_{\rm{a}}$ is the area of the sample which absorbs the electromagnetic radiation, while $A_{\rm{e}}$ is the area which emits the heat by radiation. Hence, the ratio $\frac{A_{\rm{a}}}{A_{\rm{e}}}$ equals $0.5$. $\sigma_{\rm{SB}}$ is the Stefan-Boltzmann constant, $SC$ states for Solar Constant. The thermo-optical parameters have been provided by the manufacturer of the $\rm{Upilex-S}^{\rm{\textregistered}}$ foil, the UBE company. Solar absorptance $\alpha_{\rm{S}}$ and normal emittance $\epsilon_{\rm{t}}$ are $0.093$ and $0.017$, respectively. The foil temperature as a function of distance from the Sun is represented by solid line in Fig. \ref{UBE_temperature}. Note, that the heat released by stopped protons is negligible small in comparison to the Sun's input. The red area (570 - 300 K) is the temperature range in which the bubble formation has been confirmed by the terrestrial laboratory experiments. Unfortunately, commonly used experimental procedures to estimate the maximum temperature at which the bubble formation is stopped, are not suitable for the real space conditions. The dark-red area represents temperatures at which the bubble formation has been confirmed by the first experimental findings performed at DLR. Some of the results, needed to validate the thermodynamic model are shown in Section \ref{results}. The bubble growth continues even when the probe is moving outwards from the Sun ($\ge$ 2.8 AU, grey area). Obviously, at larger distances the bubble growth slows down, since the probe is being bombarded by the smaller proton fluxes, see Eq. \ref{flux_r}.

\subsection{Thermodynamic approach to blistering process}
\label{modelA}

In the following, a thermodynamic model of bubble growth is proposed. The model is based on the assumption that the growth proceeds quasi-static i.e. during a $\rm{j^{th}}$ period of time ${\Delta}t_{\rm{j}}$ a small portion of $\rm{H_2}$-molecules, $N_{\rm{H_2,i,j}}$, is added to the $\rm{i^{th}}$ bubble and a thermodynamic equilibrium is rapidly re-established.  

For simplicity it is assumed that a single bubble is a half of a sphere with a radius of $r_{\rm{i}}$. The gas within a bubble behaves to a good approximation like an ideal gas:

\begin{equation} \label{pressure}
  p_{\rm{i}}V_{\rm{i}} = \sum_{\rm{j}}^{\rm{N}}N_{\rm{H_2, i, j}} k_{\rm{B}} T,
\end{equation}

\noindent where $p_{\rm{i}}$ is the pressure of the gas, $k_{\rm{B}}$ is the Boltzmann constant, $T$ denotes the temperature of the sample, $\rm{N}$ is the number of time steps up to a given state of bubble growth, hence the irradiation time of the sample after $N$ time steps is $N \times {\Delta}t_{\rm{j}}$.

The number of recombined H atoms, subtracted by those which diffuse from the sample out ($N_{\rm{diff,j}}$) is: 

\begin{eqnarray} \label{balance}
  N_{\rm{H,j}} &=& I_{\rm{E}} {\Delta}t_{\rm{j}} A (1 - BS) + N_{\rm{diff,j}}, \\ \nonumber
  N_{\rm{diff,j}} &=& - D_{\rm{H}}(T) \frac{\zeta_{\rm{H,j}}}{d_{\rm{PR}}(E)} \left( A - A_{\rm{b}} \right) {\Delta}t_{\rm{j}}. 
\end{eqnarray}

The term $D_{\rm{H}}(T) \frac{\zeta_{\rm{H,j}}}{d_{\rm{PR}}(E)}$ determines how much Hydrogen diffuses from the sample out (per unit area and time). Therefore, by dimension it is a flux of outflowing Hydrogen from the specimen. Its constant value results from the following fact. The model assumes that the sample is exposed to the protons having kinetic energies from eVs to 8 keV. According to the data (see, Fig. \ref{p_flux}), the magnitude of the proton flux remains almost constant for the considered energy range. Therefore, the sample is uniformly populated by the protons (H atoms after recombination) to a depth of $d_{\rm{PR}}$. The $d_{\rm{PR}}$ is calculated for the fastest protons. Therefore, there is no differentiation of Hydrogen concentration with respect to the depth.

In Eq. \ref{balance} $A$ is the area of the sample irradiated by the protons, $A_{\rm{b}}$ is the area of the sample covered by the bubbles. $BS$ is the factor of backscattered ions. If $BS$ is 1 then all of the incident ions are backscattered. If $BS$ is 0 then all of the incident ions penetrate the target. $D_{\rm{H}}(T)$ is the diffusion coefficient for H atoms in a given material, $\zeta_{\rm{H,j}}$ is the number density of H atoms which may diffuse through the lattice in the $j^{\rm{th}}$ period of time, $d_{\rm{PR}}(E)$ is the so-called projected range. It is defined as an average value of the depth to which a charged particle will penetrate in the course of slowing down to rest. This depth is measured along the initial direction of the particle, and it depends on the kinetic energy of the particle \citep{pstar}. 

The number of Hydrogen molecules added in the $j^{\rm{th}}$ period of time to the $i^{\rm{th}}$ bubble $N_{\rm{H_2, i, j}}$ is constant and given by:

\begin{eqnarray}
  N_{\rm{H_2, i, j}} & = & 0.5 G_{\rm{i}} N_{\rm{H, j}} \ \eta_{\rm{max}}(s) \ \xi, \\ \nonumber
  \sum_i G_{\rm{i}} & = & 1, \qquad \rm{dim} G = N_{\rm{B}}^{\rm{T}}. \\ \nonumber
\end{eqnarray}

\noindent Here $0.5$ denotes that a single $\rm{H_2}$ molecule consists of two H atoms, $G$ is a matrix, its role is to redistribute certain number of $\rm{H_2}$ molecules into the bubbles. The input pattern of $\rm{H_2}$ molecules into the bubbles follows experimental findings of Kamada et al. \citep{kamada}. $N_{\rm{B}}^{\rm{T}}$ is the total number of bubbles on the irradiated sample. While 100\% of protons recombine into H atoms in the metal lattice, only a part of them recombine to $\rm{H_{2}}$ molecules \citep{canham}. Hence the $\eta_{\rm{max}}(s)$ coefficient is the ratio between the number of $\rm{H_{2}}$-molecules and the H-atoms in the lattice. The $\rm{H_2}$ molecule is formed when electrons of two H atoms have anti-parallel spin $s$, otherwise the molecule cannot be created. Therefore, at most half of the H atoms can form $\rm{H_2}$ molecules, hence $\eta_{\rm{max}}(s) = 0.5$. Not all of the $\rm{H_{2}}$-molecules will merge into $\rm{H_{2}}$-clusters and finally form $\rm{H_{2}}$-bubbles. Thus, the coefficient $\xi$ denotes the ratio of the number of $\rm{H_{2}}$-molecules inside and outside the bubbles.    

The first step to estimate the radius of the $\rm{i^{th}}$ bubble is to calculate the Helmholtz free energy of the whole configuration, $F_{\rm{config}}$. Since the free energy is an additive quantity, the total free energy of bubble formation is a sum of following quantities: free energy of $\rm{H_2}$ gas inside the $\rm{i^{th}}$ bubble ($F_{\rm{gas, i}}$), of the metal surface deformation ($F_{\rm{md}, i}$) caused by the bubble growth itself, of the surface free energy ($F_{\rm{surf, i}}$) of the bubble cap, of the free energy of $\rm{H_{2}}$-molecules ($F_{\rm{H_2}}$) and of H-atoms ($ F_{\rm{H}}$) placed outside the bubbles but within the metal lattice. The Helmholtz free energy of the whole configuration described above is then: 

\begin{equation} \label{f_system}
F_{\rm{config}} = \sum_i^{\rm{N_B}}\left(F_{\rm{gas, i}} + F_{\rm{md, i}} + F_{\rm{surf, i}}\right) + F_{\rm{H_2}} + F_{\rm{H}}.
\end{equation}

\noindent The next step is to estimate the free energy of the $\rm{i^{th}}$ bubble. It consists of the free energy of the gas filled in the bubble, the free energy of metal deformation, and of the bubble cap surface free energy. 

Using the thermodynamic relation between gas pressure and its Helmholtz free energy $p = \left( \frac{\partial F}{\partial V} \right)_T$ together with the equation of state Eq. \ref{pressure}, the free energy of a gas within the $\rm{i^{th}}$ bubble is:

\begin{equation} \label{f_gas}
  F_{\rm{gas, i}} = - \sum_{\rm{j}}^{\rm{N}} N_{\rm{H_2, i, j}} k_{\rm{B}} T \ln \left( \frac{V_{\rm{max, i}}}{V_{\rm{min}}} \right),
\end{equation}

\noindent where $V_{\rm{max, i}}$ is the maximum volume of a given bubble. The model assumes that two $\rm{H_2}$ molecules form the smallest ("initial") possible bubble, its volume is denoted by $V_{\rm{min}}$. The radius of such a bubble is approximately $1.45$ Bohr radii \citep{ree}. Every bubble will crack if the pressure of the gas inside is higher than the pressure exerted by the metal deformation of the cap. The relation between the pressure of the gas, surface tension $\sigma$, and the bubble radius corresponding to $V_{\rm{max,i}}$ is \citep{lau}:  

\begin{equation} \label{cracking_condition}
  p_{\rm{gas, \ insite \ bubble}} - p_{\rm{outside \ bubble}} = \frac{2 \sigma} {r_{\rm{max, i}}}.
\end{equation} 

\noindent Since the sample is placed in vacuum, the pressure outside the bubble is set to zero. 

The free energy of metal deformation $F_{\rm{md}, i}$ caused by the gas pressure inside the bubble with radius $r_{\rm{i}}$ can be found in \citep{ll}, and is given by:

\begin{equation} \label{f_metal}
  F_{\rm{md, i}} = \frac{4\pi}{3} \frac{r_{\rm{i}}^3 (1 + \gamma)}{E_{\rm{Y}}} p_{\rm{i}}^2.
\end{equation}

\noindent Here $\gamma$ is the Poisson coefficient, i.e. ratio of transverse to axial strain of a sample material, $E_{\rm{Y}}$ is the Young's module.  

The free energy of a surface of a cap of the $\rm{i^{th}}$ bubble is given by \citep{marty}:

\begin{equation} \label{f_surface}
F_{\rm{surf, i}} = 4 \pi r_{\rm{i}}^2 \sigma(T).
\end{equation}

The Helmholtz free energy of the $\rm{H_{2}}$-molecules located at certain positions in the metal lattice but outside the bubbles is calculated in the form $F = E_{\rm{int}} - TS$. Where $E_{\rm{int}}$ is the internal energy of molecules/atoms located at certain positions in the metal lattice. Applying the statistical definition of the entropy $S$, this free energy is: 

\begin{eqnarray} \label{f_h2}
  F_{\rm{H_2}} &=& \left( N_{\rm{H_{2}}}^{\rm{T}} - \sum_{\rm{i}}^{N_{\rm{B}}^{\rm{T}}} \sum_{\rm{j}}^{\rm{N}} N_{\rm{H_{2}, i, j}} \right) \\ \nonumber
 & \times & \left[ \epsilon_{\rm{H_2}} + k_{\rm{B}}T \ln \left( \frac{N_{\rm{H_2}}^{\rm{T}} - \sum_{\rm{i}}^{N_{\rm{B}}^{\rm{T}}} \sum_{\rm{j}}^{\rm{N}} N_{\rm{H_{2}, i, j}}}{N_{\rm{0}}} \right) \right], 
\end{eqnarray}

\noindent where $N_{\rm{H_{2}}}^T$ is the total number of $\rm{H_2}$ molecules inside the sample, $\epsilon_{\rm{H_2}}$ is the binding energy of $\rm{H_{2}}$ molecule to a vacancy. A detailed derivation of the Eq. \ref{f_h2} is presented in \ref{app_f_h2}. $N_{\rm{0}}$ is the number of lattice sites, which can be expressed by:

\begin{equation} \label{lattice_sites}
  N_{\rm{0}} = N_{\rm{A}}d_{\rm{PR}}\frac{A}{M_{\rm{u}}},
\end{equation}

\noindent where $N_{\rm{A}}$ is the Avogadro's number. $M_{\rm{u}}$ is the molar mass of the sample's material.

The Helmholtz free energy of H atoms located at certain positions within the metal lattice is:

\begin{equation} \label{f_h}
  F_{\rm{H}} = \left(N_{\rm{H}}^{\rm{T}} - 2N_{\rm{H_{2}}}^{\rm{T}}\right) \left[ \epsilon_{\rm{H}} + k_{\rm{B}}T \ln \left( \frac{N_{\rm{H}}^{\rm{T}} - 2N_{\rm{H_{2}}}^{\rm{T}}}{N_{\rm{0}}}\right) \right],
\end{equation}

\noindent where $\epsilon_{\rm{H}}$ is the migration energy of the H atom in the metal lattice, and $N_{\rm{H}}^{\rm{T}}$ is the total number of H atoms in the sample. A detailed derivation of the Eq. \ref{f_h} can be found in \ref{app_f_h}. 

Since now each term of Eq. \ref{f_system} is determined, the next step is to estimate the radius $r_{\rm{i}}$ of the $\rm{i^{th}}$ bubble at given time $t$. This will be achieved by assuming that the process of bubble growth proceeds quasi-static in thermodynamic equilibrium: 

\begin{equation} \label{condition}
  \frac{\partial F_{\rm{config,i}}}{\partial N_{\rm{H_2, i, j}}} = 0.
\end{equation}

\noindent This condition leads to the following fifth order equation for $r_{\rm{i}}$:

\footnotesize
\begin{eqnarray} \label{bfc} 
 &8 \pi \Xi_{\rm{i,j}} \sigma(T) r_{\rm{i}}^5 - H_{\rm{i}}r_{\rm{i}}^4 + \frac{3}{\pi} \frac{1+\gamma}{E_{\rm{Y}}} \left(\sum_{\rm{j}}^{\rm{N}} N_{\rm{H_2, i, j}} \right) k_{\rm{B}}^2 T^2 \\ \nonumber
& \times \left[ 2 N r_{\rm{i}} - 3 \Xi_{\rm{i,j}} \sum_{\rm{j}}^{\rm{N}} N_{\rm{H_2, i, j}} \right] = 0,
\end{eqnarray}
\normalsize

\noindent $\Xi_{\rm{i,j}}$ is defined below in Eq. \ref{dyn_model}, $H_{\rm{i}}$ denotes the abbreviation:

\begin{equation}
H_{\rm{i}} = - \frac{\partial F_{\rm{gas, i}}}{\partial N_{\rm{H_2, i, j}}} - \frac{\partial F_{\rm{H}}}{\partial N_{\rm{H_2, i, j}}} - \frac{\partial F_{\rm{H_2}}}{\partial N_{\rm{H_2, i, j}}}. 
\end{equation}

\noindent The derivatives of the Helmholtz free energy, of the gas inside the $\rm{i^{th}}$ bubble, of metal deformation caused by the bubble, of the $\rm{i^{th}}$ bubble cap surface, of $\rm{H_2}$ molecules, and of H atoms with respect to the number of $\rm{H_2}$ molecules added at the $\rm{j^{th}}$ time step to the $\rm{i^{th}}$ bubble are presented in \ref{a_f_system}.

A realistic model of bubble radius growth, $\Xi(i,j)$, can be estimated by following Gedankenexperiment. Obviously at the beginning of the bubble growth process, the differential increase of the bubble radius is higher than at its end. It is implied, that the number of $\rm{H_2}$ molecules in the system is conserved and at each time step one of them merge into a bubble. After $\Delta{t}$ the bubble consists of $2\rm{H_2}$ molecules, hence the number of molecules increases by $50\%$. At the time $2\Delta{t}$ the bubble consists of $3\rm{H_2}$ molecules, hence the number increase is now $33.3\%$, and so on. Therefore $\Xi$ is: 

\begin{equation} \label{dyn_model}
  \Xi_{\rm{i,j}} = \frac{\Delta r_{\rm{i}}}{\Delta N_{\rm{H_2,i,j}}} = j^{\alpha} r_{\rm{i, 0}}, \qquad \alpha = \frac{1}{3} 
\end{equation}  

\noindent The exponent $\alpha$ is a model parameter of the bubble growth. The value $\frac{1}{3}$ corresponds to the Gedankenexperiment presented above. However, the true value of the $\alpha$ parameter differs from that. In the process of bubble growth, particles (the Hydrogen) are added to the system i.e. the probe is permanently irradiated by the protons, they penetrate the target and recombine to the Hydrogen. On the other hand, both, due to the diffusion process and bubble cracking, some Hydrogen atoms leave the system. Therefore, the number of Hydrogen atoms in the system is not conserved. Hence, a series of experiments have been performed to estimate a realistic $\alpha$ parameter, results are presented in Section \ref{results}.

\subsection{Reflectivity of a metallic foil covered with bubbles} \label{reflectivity}

\begin{figure}[!h]
  \begin{center}
    \includegraphics[width=0.45\textwidth]{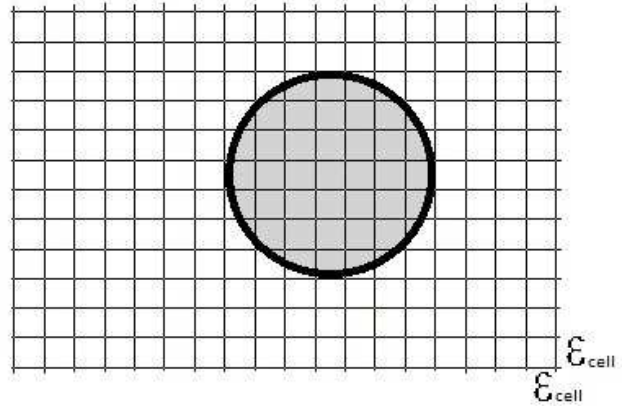}
  \end{center}
  \caption{A fraction of the foil with one spherical bubble is shown. The size of a single cell of the grid is $\epsilon_{\rm{cell}} \times \epsilon_{\rm{cell}}$.}
  \label{cells}
\end{figure}

The momentum transfer of a photon to an ideal reflecting surface is given by $\Delta{q}=2q\cos \theta$, where the factor 2 is just in accordance with specular reflectivity. Certainly, the surface quality will suffer during the irradiation with protons from progressing bubble formation. At time $t=0$ the foil has not been exposed to the electromagnetic radiation and/or charged particles, and is considered to be a perfect mirror with the reflectivity of $R=1$. It means that all of the incident light rays are reflected perfectly, no light ray is absorbed or diffusively reflected by the target. Later, when the foil has been irradiated by a flux of protons and molecular Hydrogen bubbles have been formed on its surface, the reflectivity of the degraded foil will be reduced. This deterioration is calculated in the following way: the foil is covered by a grid with a fixed single cell size of $\epsilon_{\rm{cell}} \times \epsilon_{\rm{cell}}$, see Fig. \ref{cells}. The reflectivity of a single cell is by definition $\frac{\Delta q}{\Delta q_{\rm{max}}}$, where ${\Delta}q$ is momentum transfer of a photon to the $\rm{i^{th}}$ cell of the degraded foil, while ${\Delta}q_{\rm{max,i}}$ is the momentum transfer of a photon to the $\rm{i^{th}}$ cell of a perfect mirror. 

Therefore, taking into account all cells, one has:

\begin{equation} \label{reflectivity_common}
\Delta R = \frac{\sum_{\rm{i}}^{\rm{N_{cell}}} \Delta q_{\rm{i}} }{ \sum_{\rm{i}}^{\rm{N_{cell}}} \Delta q_{\rm{max, i}} }.
\end{equation}

\noindent Here $N_{\rm{cell}}$ is the number of cells. The path of photons is directed parallel to the foil surface normal. Therefore, at time $t=0$ the foil was a perfect mirror without surface imperfections and $\theta_{\rm{i}}=0$. Later, when the surface is populated with bubbles, $\theta_{\rm{i}}$ will vary between $0^o$ and $90^o$. Thus, Eq. \ref{reflectivity_common} reduces to: 

 \begin{equation}
\Delta R = \frac{\sum_{\rm{i}}^{\rm{N_{cell}}} 2q \cos \theta_{\rm{i}} }{ N_{\rm{cell}} \times 2q } =  \frac{\sum_{\rm{i}}^{\rm{N_{cell}}} \cos \theta_{\rm{i}} }{ N_{\rm{cell}} }.
\end{equation}

\section{Results} 
\label{results}

\begin{figure}
\centering
\includegraphics[width=0.35\textwidth]{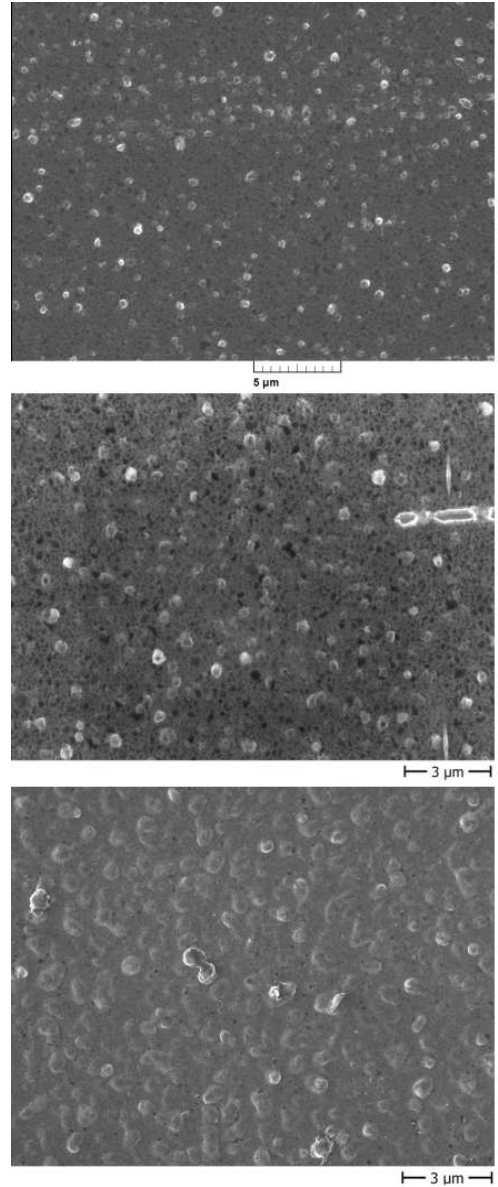}
\caption{Electron microscope pictures of probes A1 (top), A2 (middle), and A3 (bottom).}
\label{dose_dep}
\end{figure}

A following set of experiments were performed. Three probes (A1, A2, and A3) were exposed to a flux of 2.5 keV protons, each one with longer irradiation time, see Table \ref{test_parameters}, where $t_{\rm{S}}$ is a number of days in space until a probe will collect a given dose of protons. Results are shown in Fig. \ref{dose_dep}. From top to bottom, the pictures correspond to the probes A1, A2, and A3, respectively. Average sizes of bubbles have been estimated to $0.18 \ \pm 0.05$ $\rm{\mu}$m, $0.19 \ \pm 0.05$ $\rm{\mu}$m, and $0.2 \ \pm 0.05$ $\rm{\mu}$m for probe A1, A2, and A3, respectively. There is a strict correlation between a dose of protons and the average bubble size for a given population. The higher the proton dose, the larger the bubble sizes. Examining the electron microscope pictures, the surface density of bubbles has been estimated to $\sim 10^8$ $\rm{cm^{-2}}$.

We have confirmed formation of molecular Hydrogen bubbles by a number of experiments. We have varied protons flux, energy, and specimens' temperature. Bubbles were observed for specific parameters range. Fig. \ref{UBE_temperature} represents our findings.

The here used three experimental findings are the minimum number of specimens necessary to validate the model.

\begin{table}[!ht]
\centering
\caption{Test parameters}
\vspace{3.pt}
\scalebox{0.85}{\begin{tabular}{ccccc}
\hline
Probe symbol & $T$ [K] & $E$ [keV] & $D$ [$\rm{p^+ \ cm^{-2}}$] & $t_{\rm{S}}$ [days] \\
\hline
A1 & 323.0 & 2.5 & $7.8 \times 10^{17}$ & 4.8\\
A2 & 323.0 & 2.5 & $8.2 \times 10^{17}$ & 5.0\\
A3 & 323.0 & 2.5 & $1.3 \times 10^{18}$ & 7.9\\
\hline
\end{tabular}}
\label{test_parameters}
\end{table}

\subsection{Bubble growth law} \label{bgl}

Bubble growth in blistering process  is dominated by diffusion of hydrogen atoms into the bubbles. A similar problem of the bubble growth (Ostwald ripening) was widely discussed in literature \citep[e.g.][]{chai,jones,lif,lif2,rut,wagner}. It has been found the power law $\bar{r}(t) \sim t^{\frac{1}{3}}$  for the average bubble radius $\bar{r}$ in the systems with the total mass being a conserved quantity. This type of the systems exhibits diffusion limited growth and is well described in terms of the Lifshitz-Slyozov-Wagner (LSW) theory \citep{lif,wagner}. The Hydrogen bubbles follow the diffusion limited growth and the question can be raised whether the LSW theory applies to them.  In this case the growth rate of the bubble average radius $\bar{r}(t)$ could be represented by the following phenomenological equation:

\begin{equation}
 \frac{d\bar{r}(t)}{dt} = \frac{ D_{\rm{H}} }{ \bar{r}(t) } \left[ \Delta(t) - \frac{\alpha_{\rm{C}}}{ \bar{r}(t)}  \right] \,\,,
\label{eqLS}
\end{equation}

\noindent similarly as in \citep{lif}, where $D_{\rm{H}}$ is the diffusion coefficient of Hydrogen atoms,

\begin{equation}
  \alpha_{\rm{C}} = \frac{2 \sigma V_{\rm{m}} C_{\rm{\infty}}}{k_{\rm{B}}T},
\end{equation}

\noindent
is a parameter which depends on the surface tension $\sigma$ of a bubble, the Hydrogen atom volume $V_{\rm{m}}$, the concentration $C_{\rm{\infty}}$ of Hydrogen atoms far beyond the bubble and temperature T, $\Delta(t)$ is the difference of concentration of Hydrogen atoms at the bubble boundary and $C_{\rm{\infty}}$. In the LSW approach, the total mass of a system is a conserved quantity, in consequence $\frac{\bar{r}(t) \Delta(t)}{\alpha_{\rm{C}}} = \rm{const}$, and the solution of Eq. \ref{eqLS} takes the form of $\bar{r}(t) \sim t^{1/3}$. 

\begin{figure}
\centering
\includegraphics[width=0.45\textwidth]{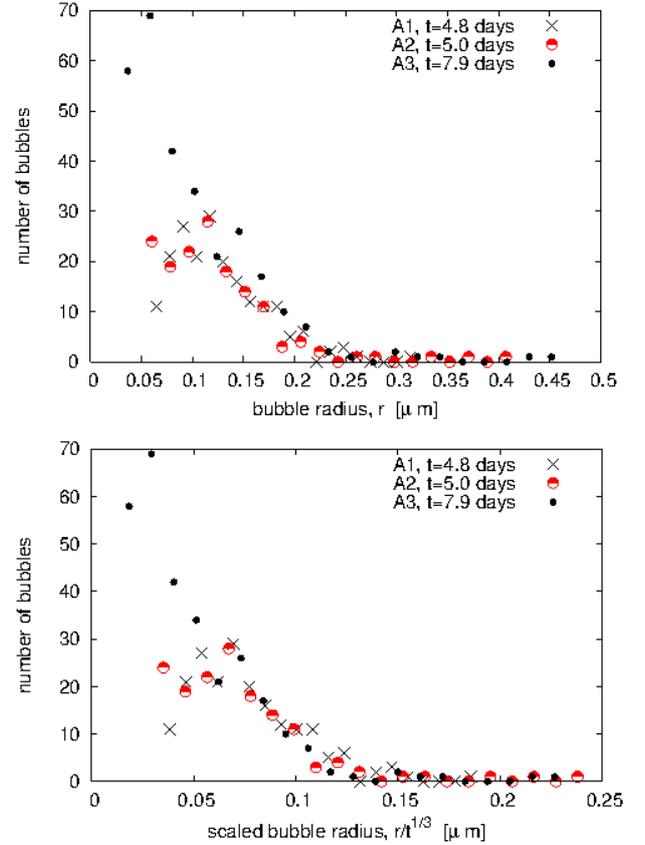}
\caption{Top plot, distribution of bubble radius $r$ for bubbles in Fig. \ref{dose_dep}, respectively for probe A1, A2 and A3. Bottom plot, the same as top plot but for the scaled radius, $r/t^{1/3}$.}
\label{models_distribution_data}
\end{figure}

The three probes $A_1$, $A_2$, $A_3$ from the pictures in Fig. \ref{dose_dep} were exposed to a constant proton flux. Taking into account that the amount of the Hydrogen atoms produced by the proton flux is related directly to the existing electronic structure of the irradiated probes as well as  that the thickness of the superficial layer including the molecular Hydrogen bubbles is relatively constant in size (it only shifts deeper into the sample after the superficial bubbles break) then the system of the growing bubbles resembles the total mass conservation system. Hence, it could be expected that at least the largest bubbles 
which appear on the irradiated
probes surface follow the growth law $t^{1/3}$ as well as they undergo coalescence phenomena. The latter property is evident, e.g. for
probe $A_3$ in Fig. \ref{models_distribution_data}. Concerning the question of the possible existence of the $t^{1/3}$ scaling, 
the average bubble radius, respectively  for the probe $A_1$, $A_2$, $A_3$ (Table 2) has been divided by the duration 
of the proton flux exposure in the power of $1/3$, i.e., $4.8^{1/3}$, $5^{1/3}$ and $7.9^{1/3}$. All scaled radii take the same value close to $0.1$. In addition, in Fig. \ref{models_distribution_data}, the distribution of 
the bubble radii, $r/t^{1/3}$ of the growing  bubbles has been plotted. It can be observed that the largest 
bubbles 
seem to follow the $t^{1/3}$ power growth law  because their scaled distributions coincide.  
The left hand side of the bubble radius distribution is influenced by the bubbles
which were formed  at later time moments  and it is the reason for the strong deviation of their size 
from the $t^{1/3}$ power law. It can be concluded from the surface analysis of the probes
$A_1$, $A_2$ and $A_3$ 
that their degradation after they were exposed to proton flux
is not faster than it is predicted by LSW theory. However, it is necessary further investigation on the problem 
both experimentally and theoretically.

\subsection{Validation of the thermodynamic model} \label{TH_validation}

For numerical simulation a $10 {\mu}m \times 10 {\mu}m$ foil was specified. That choice allows to simulate a smaller number of bubbles, i.e. it decreases the computation time of the simulation. It implies also an important assumption that surface arrangement of the bubbles is isotropic i.e. any $10 {\mu}m \times 10 {\mu}m$ area of the irradiated sample is indistinguishable. Table \ref{model_parameters} collects all of the model parameters used in the simulation. The first set of parameters characterize mechanical and thermo-optical properties of vacuum deposited Aluminum on UBS's $\rm{Upilex-S^{\textregistered}}$ foil. Second set specifies values of the parameters which have been used to fit the model to the experimental data presented here.

\begin{figure}
\centering
\includegraphics[width=0.5\textwidth]{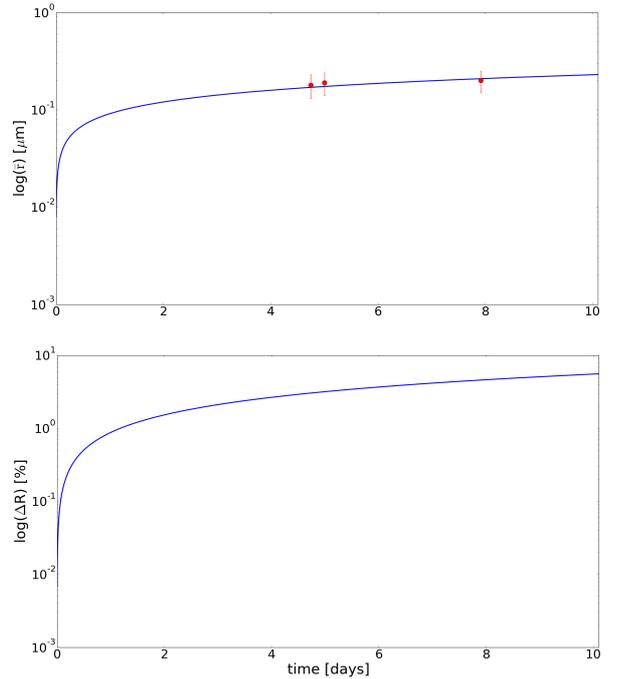}
\caption{Time evolution of an average bubble radius from the population (top plot), specular reflectivity decrease due to bubble growth (bottom plot).}
\label{model}
\end{figure}

\begin{figure}
\centering
\includegraphics[width=0.5\textwidth]{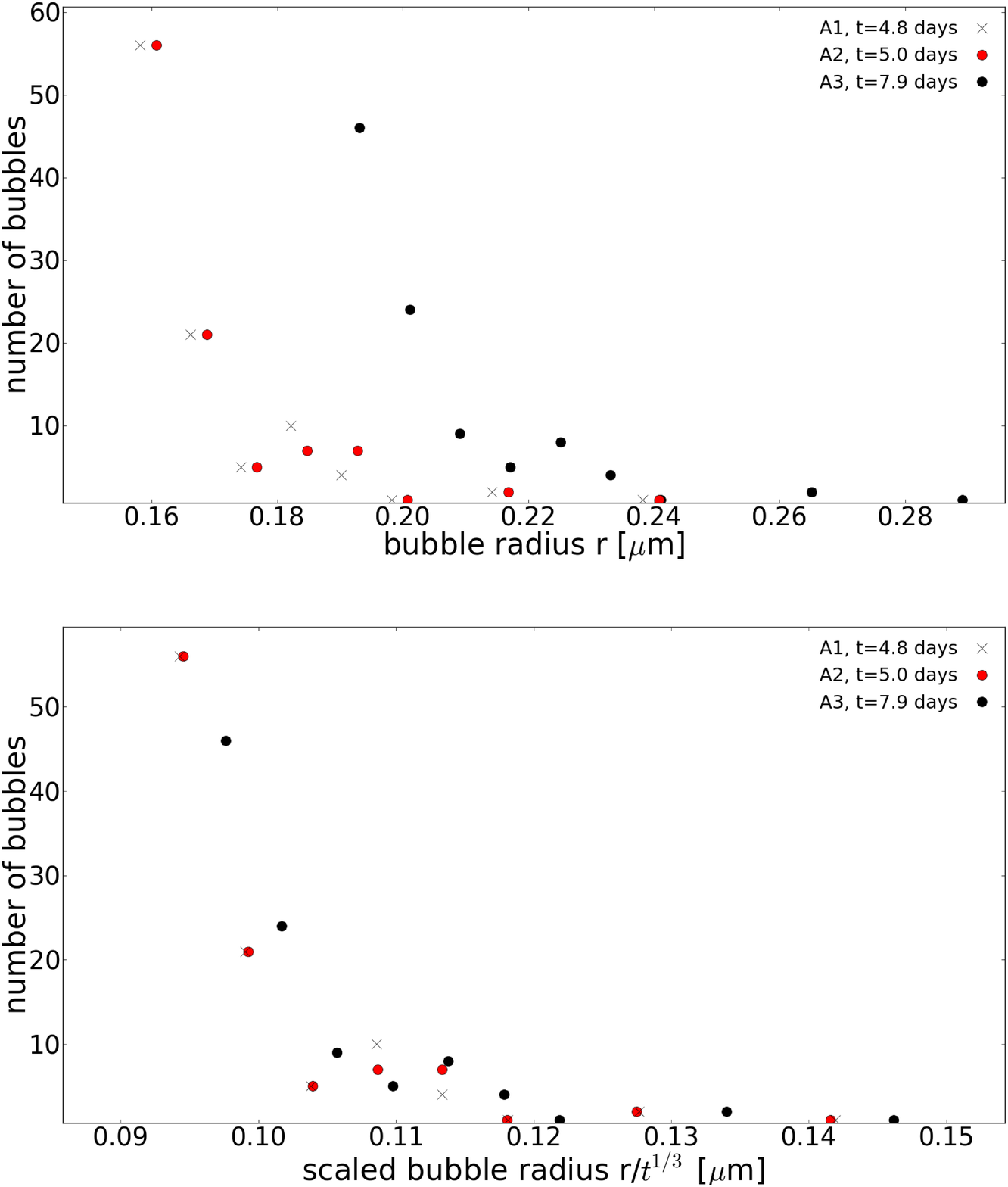}
\caption{Bubble size distribution at a $100$ $\rm{\mu m^2}$ sample at three different time steps: $4.8$, $5.0$, and $7.9$ $t_{\rm{S}}$.}
\label{models_distribution}
\end{figure}

To fit the proper gradient of bubble growth, the $\alpha$ parameter was set to $0.6$, see Eq. \ref{dyn_model}. Comparison of the average bubble size of the experimental and numerical findings are drawn in the top plot of the Fig. \ref{model}. The $\xi$ parameter was set to $0.98$. It determines the height of the curve. A decrease of the specular reflectivity of the foil as a function of time is shown in the bottom plot of the Fig. \ref{model}. The decrease of the reflectivity is $3.0$, $3.2$, and $4.6$ \% in comparison to the non-irradiated foil for $t_{\rm{S}} = 4.75$, $5.0$, and $7.9$ days, respectively. Clearly, the larger the bubble sizes, the larger the specular reflectivity decrease ${\Delta}R$ in comparison to the non-irradiated foil. A distribution of the bubbles at three different time steps: $t_{\rm{S}} = 4.8$, $5.0$, and $7.9$ days is shown in Fig. \ref{models_distribution}. During time, the probe collects higher dose of protons, therefore, the distribution drifts i.e. size of bubbles increases. At time $t_{\rm{S}} = 7.9$ days most of the bubbles have sizes in range $0.19$ to $0.24$ $\rm{\mu}$m, there are only a few which have sizes larger than $0.25$ $\rm{\mu}$m.

\begin{table}[!ht]
\centering
\caption{Model parameters}
\vspace{3.pt}
\scalebox{1.0}{\begin{tabular}{cl}
\hline
Symbol & Value \\
\hline
$\varrho$ & $2.7$ [$\rm{g \ cm^{-3}}$] \\
$M_{\rm{u}}$ & $26.98$ [$\rm{g \ mol^{-1}}$] \\
$E_{\rm{Y}}$ & $69 \times 10^{10}$ [$\rm{dyn \ cm^{-2}}$] \\
$\gamma$ & $0.33$ \\
$\epsilon_{\rm{H}}$ & 0.52 [eV] \citep{lind} \\
$\epsilon_{\rm{H_2}}$ & 0.06 [eV] \citep{lu} \\
$\alpha_{\rm{S}}$ & 0.093 \\
$\epsilon_{\rm{t}}$ & 0.017 \\
\hline
$BS$ & 0.02 \citep{srim} \\
$A$ & 100 [$\rm{{\mu}m^2}$] \\
$T$ & 323 [K] \\
$\eta_{\rm{max}}(s)$ & 0.5 \\
$\xi$ & 0.98 \\
$\alpha$ & 0.6 \\
$N_{\rm{B}}$ & $10^8$ [$\rm{cm^{-2}}$] \\
\hline
\end{tabular}}
\label{model_parameters}
\end{table}

\section{Conclusions}
\label{conclusions}

It has been proven that thermodynamic model is a flexible tool to simulate and to reproduce the real growth of the molecular Hydrogen bubbles. However, the estimated $\alpha$ and $\xi$ parameters are applicable only for the here presented experimental findings. These parameters depend on type and temperature of the irradiated material. Therefore, change of the material type and the experimental conditions requires an update of the model parameters.

The time evolution of decrease of the specular reflectivity ${\Delta}R$ is a model prediction. It is highly required to perform experimental confirmation of that findings, since the real reflectivity decrease can differ from that estimated here.

For sure, the here presented experimental findings show that physical parameters of the metal surface, e.g. roughness, reflectivity, are significantly changed under conditions as prevalent in the interplanetary space. Therefore, statement present in the ECSS standard \citep[][p. 51]{ecss-71A} that metals do not suffer from space-environment is not longer valid.

The thermodynamic model requires further improvements. The considered aging factor, the solar protons, are not the only one which can influence the bubble growth process. The solar wind is also essentially made up of electrons and small proportion of heavier ions \citep{sw}. Additionally, electromagnetic radiation in the lower wavelength range has also to be taken into consideration. These degradation factors can slow down the bubble growth. The growth deceleration can be explained as follows. Hydrogen molecule dissociate at the energy of $4.5$ eV \citep{balak}. The dissociation may be caused by the UV-light with the wavelengths $\le 274$ nm. The $\rm{H_2}$ gas within the bubbles can then be partially dissociated, and H atoms can diffuse easily through the bubble caps. As a result the bubble growth process may slow down. The deceleration can be strengthen by heavier ions generated by the Sun e.g. $\alpha$-particles. Their diameter is much larger than that of protons or electrons, hence, collisions between the $\rm{H_2}$ molecules and the $\alpha$-particles within the bubbles can additionally increase the dissociation efficiency.

The present condition for the bubble crack mechanism, Eq. \ref{cracking_condition}, assumes that the pressure outside the bubbles is negligible small. Under the real space conditions the electromagnetic radiation will exert a pressure on the caps, hence, their sizes may be smaller. On the other hand, bubble caps loose thermal contact with the base material and they become overheated \citep{astrelin}. As a consequence the caps can brake and the $\rm{H_2}$ gas can be released. That aspect of the blistering process needs to be examined.

By these reasons further experimental studies are planned, e.g. the mentioned influence of the UV-light on the bubble growth dynamics. 

\section{Acknowledgements}

We would like to express our special thanks to dr. Herbert Juling who performed electron microscope measurements and Mrs. Shiho Otsu from UBE Industries Ltd. who provides us the foils. 

\appendix
\section{Bubble formation}
\subsection{Helmholtz free energy of $\rm{H_2}$ molecules placed in certain positions in the metal lattice} \label{app_f_h2}

To get the relation of the Helmholtz free energy of the $\rm{H_{2}}$ molecules located at certain positions in the sample but outside the bubbles, the definition of the free energy is used:

\begin{equation} \label{free_enery}
  F = E_{\rm{int}} - TS, \qquad S = k_{\rm{B}} \ln \Omega,
\end{equation} 

\noindent here $E_{\rm{int}}$ is the internal energy of the $\rm{H_{2}}$ molecules located at certain positions in the metal lattice, $S$ denotes the entropy, while $\Omega$ represents the number of ways in which the $\rm{H_{2}}$ molecules can be arranged on the $N_{\rm{0}}$ lattice sites (Eq. \ref{lattice_sites}). 

The number of $\rm{H_{2}}$ molecules located outside the bubbles at certain positions in the metal lattice is $N_{\rm{H_{2}}}^{\rm{T}} - \sum_{\rm{i}}^{N_{\rm{B}}^{\rm{T}}} \sum_{\rm{j}}^{N} N_{\rm{H_{2}, i, j}}$. Where $N_{\rm{H_{2}}}^{\rm{T}}$ is the total number of $\rm{H_{2}}$ molecules in the sample, while $\sum_{\rm{i}}^{N_{\rm{B}}^{\rm{T}}} \sum_{\rm{j}}^{N} N_{\rm{H_{2}, i, j}}$ is the total number of $\rm{H_{2}}$ molecules within all bubbles. The summation over the number of time steps $j$ counts the number of $\rm{H_{2}}$ molecules in the $\rm{i^{th}}$ bubble. The second summation over number of bubble $\rm{i^{th}}$ counts the number of $\rm{H_{2}}$ molecules within all bubbles. Therefore, performing the subtraction one gets the total number of $\rm{H_2}$ molecules outside all bubbles, located at certain positions in the metal lattice. The entropy $k_{\rm{B}} \ln \Omega$ of the collection of the $\rm{H_{2}}$ molecules placed on a lattice sites is \citep{seitz}:

\scriptsize
\begin{align} \nonumber
S& = k_{\rm{B}}\ln \frac{N_{\rm{0}} !}{\left(N_{\rm{H_{2}}}^{\rm{T}} - \sum_{\rm{i}}^{N_{\rm{B}}^{\rm{T}}} \sum_{\rm{j}}^{\rm{N}} N_{\rm{H_{2}, i, j}}\right)!\left[N_{\rm{0}} - \left(N_{\rm{H_{2}}}^{\rm{T}} - \sum_{\rm{i}}^{N_{\rm{B}}^{\rm{T}}} \sum_{\rm{j}}^{\rm{N}} N_{\rm{H_{2}, i, j}}\right) \right]!}, \\ 
&\cong - k_{\rm{B}} \left(N_{\rm{H_{2}}}^{\rm{T}} - \sum_{\rm{i}}^{N_{\rm{B}}^{\rm{T}}} \sum_{\rm{j}}^{\rm{N}} N_{\rm{H_{2}, i, j}}\right) \ln \left[\frac{N_{\rm{H_{2}}}^{\rm{T}} - \sum_{\rm{i}}^{N_{\rm{B}}^{\rm{T}}} \sum_{\rm{j}}^{\rm{N}} N_{\rm{H_{2}, i, j}}}{N_{\rm{0}}}\right].
\end{align}   
\normalsize

\noindent The internal energy $E_{\rm{int}}$ is given by the following relation:

\begin{equation}
  E_{\rm{int}} = \epsilon_{\rm{H_2}} \left(N_{\rm{H_{2}}}^{\rm{T}} - \sum_{\rm{i}}^{N_{\rm{B}}^{\rm{T}}} \sum_{\rm{j}}^{\rm{N}} N_{\rm{H_{2}, i, j}}\right),
\end{equation}

\noindent where the $\epsilon_{\rm{H_2}}$ is the binding energy of the $\rm{H_{2}}$ molecule to a vacancy \citep{lu}. The internal energy of $\rm{H_{2}}$ molecules located in the metal lattice sites is a product of the binding energy of a single $\rm{H_{2}}$ molecule and the number of molecules. 

The Helmholtz free energy of the $\rm{H_{2}}$ molecules located outside the bubbles at certain positions in the metal lattice is then:

\begin{eqnarray} \label{a_f_h2}
  F_{\rm{H_2}} &=& \left( N_{\rm{H_{2}}}^{\rm{T}} - \sum_{\rm{i}}^{N_{\rm{B}}^{\rm{T}}} \sum_{\rm{j}}^{\rm{N}} N_{\rm{H_{2}, i, j}} \right)\\ \nonumber
 &\times& \left[ \epsilon_{\rm{H_2}} + k_{\rm{B}}T \ln \left(\frac{N_{\rm{H_{2}}}^{\rm{T}} - \sum_{\rm{i}}^{N_{\rm{B}}^{\rm{T}}} \sum_{\rm{j}}^{\rm{N}} N_{\rm{H_{2}, i, j}}}{N_{\rm{0}}} \right) \right].
\end{eqnarray}

\subsection{Helmholtz free energy of H atoms in the sample} \label{app_f_h}

The number of H atoms in the sample and outside the bubbles is:

\begin{equation} \label{a_n_h}
  N_{\rm{H}} = N_{\rm{H}}^{\rm{T}} - 2 \left(\sum_{\rm{i}}^{\rm{N_B}} \sum_{\rm{j}}^{\rm{N}} N_{\rm{H_2, i, j}} + N_{\rm{H_2}}^{\rm{out. \ bubbles}}\right),
\end{equation}

\noindent where $N_{\rm{H}}^{\rm{T}}$ is the total number of H atoms in the sample, so the number counts all of the incident Hydrogen ions which have recombined into Hydrogen atoms. Some of the Hydrogen atoms have recombined to $\rm{H_2}$ molecules and some of the molecules are forming the bubbles. Hence to get the number of H atoms located on the lattice sites one has to subtract the total number of Hydrogen atoms $\rm{N_H^T}$ and those Hydrogen atoms which build $\rm{H_2}$ clusters and $\rm{H_2}$ bubbles. The reason of the factor $2$ is that a single $\rm{H_{2}}$ molecule consists of two H atoms.   

The procedure to estimate the Helmholtz free energy of H atoms in the sample is the same as in Eq. \ref{a_f_h2}. Hence the term is:

\begin{equation} \label{a_f_h}
  F_{\rm{H}} = N_{\rm{H}} \left(\epsilon_{\rm{H}} + k_{\rm{B}}T \ln \frac{N_{\rm{H}}}{N_{\rm{0}}} \right),
\end{equation}

\noindent where $\epsilon_{\rm{H}}$ is the migration energy of the H atom in the metal lattice. The migration energy is defined as the minimum energy which has to be added to the H atom in order to remove it from the lattice site.  

\subsection{The derivatives of Helmholtz free energy of: gas of the $i^{\rm{th}}$ bubble, metal deformation caused by the bubble, surface of the $\rm{i^{th}}$ bubble cap, $\rm{H_2}$ molecules, and $\rm{H}$ atoms located outside the bubbles.} \label{a_f_system}

The equlibrium condition of the process of the $\rm{i^{th}}$ bubble growth is:

\begin{equation} \label{condition}
  \frac{\partial F_{\rm{config,i}}}{\partial N_{\rm{H_2, i, j}}} = 0,
\end{equation} 

The assumption is fulfilled when the time scale of the bubble growth is longer than the time scale of the formation of a $\rm{H_2}$ molecule out of two H atoms. The thermodynamic equilibrium is rapidly re-established after merging a $\rm{H_2}$ molecule to a given $\rm{i^{th}}$ bubble during a given time step. Condition \ref{condition} can be written as a sum:

\begin{equation}
  \frac{\partial F_{\rm{gas, i}}}{\partial N_{\rm{H_2, i, j}}} + \frac{\partial F_{\rm{md, i}}}{\partial N_{\rm{H_2, i, j}}} + \frac{\partial F_{\rm{surf, i}}}{\partial N_{\rm{H_2, i, j}}} + \frac{\partial F_{\rm{H_2}}}{\partial N_{\rm{H_2, i, j}}} + \frac{\partial F_{\rm{H}}}{\partial N_{\rm{H_2, i, j}}} = 0.
\end{equation}

\noindent Derivatives of the free energy of the gas in the $\rm{i^{th}}$ bubble, of metal deformation caused by the bubble, surface of the $\rm{i^{th}}$ bubble cap, and of $\rm{H_2}$ molecules and H atoms located on the lattice sites with respect to the number of $\rm{H_2}$ molecules that merge on each time step to a bubble, will be calculated separately. By use of the Helmholtz free energy of the gas, Eq. \ref{f_gas}, the derivative is:

\begin{equation}
  \left. \frac{\partial F_{\rm{gas, i}}}{\partial N_{\rm{H_2, i, j}}}\right|_{\rm{N}} = - N k_{\rm{B}} T \ln \left( \frac{V_{\rm{max}, i}}{V_{\rm{min}}} \right) - \frac{3}{2} N k_{\rm{B}} T.
\end{equation}

 \noindent The free energy of a metal deformation caused by expanding $\rm{i^{th}}$ bubble is given by the Eq. \ref{f_metal}, hence its derivative is:

\begin{eqnarray}
  \left. \frac{\partial F_{\rm{md, i}}}{\partial N_{\rm{H_2, i, j}}}\right|_{\rm{N}} &=& \frac{3}{\pi} \frac{1+\gamma}{E_{\rm{Y}}} k_{\rm{B}}^2 T^2 \sum_{\rm{j}}^{\rm{N}} N_{\rm{H_2, i, j}} \\ \nonumber
 &\times& \left[2r_{\rm{i}}^{-3}N  - 3 r_{\rm{i}}^{-4} \frac{\partial r_{\rm{i}}}{\partial N_{\rm{H_2, i, j}}} \sum_{\rm{j}}^{\rm{N}} N_{\rm{H_2, i, j}} \right].
\end{eqnarray}

\noindent The free energy of a surface of the $\rm{i^{th}}$ bubble cap is given by the Eq. \ref{f_surface}. The corresponding derivative is given by:

\begin{equation}
\left. \frac{\partial F_{\rm{surf, i}}}{\partial N_{\rm{H_2, i, j}}}\right|_{\rm{N}} = 8 \pi r_{\rm{i}} \frac{\partial r_{\rm{i}}}{\partial N_{\rm{H_2, i, j}}} \sigma(T).
\end{equation}

\noindent The derivative of the Helmholtz free energy of the $\rm{H_{2}}$ molecules (Eq. \ref{a_f_h2}) located outside the bubbles at certain positions in the metal lattice is:

\scriptsize
\begin{equation}
  \left. \frac{\partial F_{\rm{H_2}}}{\partial N_{\rm{H_2, i, j}}}\right|_{\rm{N}} = - N N_{\rm{B}}^{\rm{T}} \left\{ \epsilon_{\rm{H_2}} + k_{\rm{B}} T \left[ 1 + \ln \left(\frac{N_{\rm{H_2}}^{\rm{T}} - \sum_{\rm{i}}^{\rm{N_{\rm{B}}^{\rm{T}}}} \sum_{\rm{j}}^{\rm{N}} N_{\rm{H_2, i, j}}}{N_{\rm{0}}} \right)\right]\right\}.
\end{equation} 
\normalsize

\noindent The derivative of the Helmholtz free energy of the H atoms (Eqs. \ref{a_f_h}) located at certain positions in the metal lattice is:

\begin{equation}
  \left. \frac{\partial F_{\rm{H}}}{\partial N_{\rm{H_2, i, j}}}\right|_{\rm{N}} = -2 N N_{\rm{B}}^{\rm{T}} \left[ \epsilon_{\rm{H}} + k_{\rm{B}} T \left( 1 + \ln \frac{N_{\rm{H}}^{\rm{T}} - 2N_{\rm{H_2}}^{\rm{T}}}{N_{\rm{0}}} \right) \right].
\end{equation}

\end{document}